\documentstyle[amsfonts,amssymb,11pt,epsfig]{article}

\makeatletter
\@addtoreset{equation}{section}

\newcommand{\R}{{\Bbb R}}
\newcommand{\C}{{\Bbb C}}

\def\SL{{\rm SL}}
\def\H{{{\bf H}^3}}

\newcommand{\be}{\begin{eqnarray}} 
\newcommand{\ee}{\end{eqnarray}}

\begin{document}

\begin{titlepage}

\thispagestyle{empty}

\title{$\Lambda<0$ Quantum Gravity in 2+1 Dimensions I: \\
Quantum States and Stringy S-Matrix}

\author{
{\bf Kirill Krasnov}\thanks{{\tt 
krasnov@cosmic.physics.ucsb.edu}}
\\
{\it Department of Physics}\\
{\it University of California, Santa Barbara, CA 93106}}

\date{\normalsize April, 2001}
\maketitle

\begin{abstract}
\normalsize We consider the theory of pure gravity in 2+1 dimensions,
with negative cosmological constant. The theory contains simple matter
in the form of point particles; the later are classically described as 
lines of conical singularities. We propose a formalism in which quantum
amplitudes for process involving black holes and point particles are 
obtained as Liouville field theory (LFT) correlation functions on Riemann 
surfaces $X$. Point particles are described by LFT vertex operators, black 
holes (asymptotic regions) are in correspondence with boundaries of $X$.
We analyze two examples: the amplitude for emission of a particle by the 
BTZ black hole, and the amplitude of black hole creation by two point 
particles. We then define an inner product between quantum states. The 
value of this inner product can be interpreted as the amplitude for one set 
of point particles to go into another set producing black holes. The full 
particle S-matrix is then given by the sum of all such amplitudes. This 
S-matrix is that of a non-critical string theory, with the world-sheet CFT 
being essentially the Liouville theory. $\Lambda<0$ quantum gravity 
in 2+1 dimensions is thus a string theory.
\end{abstract}

\end{titlepage}

\section{Introduction}
\label{sec:intr}

The present paper develops a quantum
theory of gravity in (2+1)-dimensions, with negative cosmological constant.
We consider pure gravity, that is a theory of only the metric. The
only type of matter we allow is a point particle, which is classically
described as a line of conical singularities. This is a purely geometrical
description, with no extra fields describing particles present. This
should be contrasted to by now standard setting of AdS${}_3$/CFT
correspondence of string theory, where the 3d theory that
appears contains, in addition to the metric, a large number of other 
fields. 

The main motivation for studying this seemingly much too simple theory 
is that, despite its deceiving simplicity, it contains all interesting
phenomena one can expect of a gravity theory. The theory is rich enough
in that it contains both black holes \cite{BTZ} and matter in the form of
point particles \cite{Jackiw}. The later can collide and form the former, 
as was first demonstrated by Matschull in \cite{Matsch}, see also 
\cite{Matsch-Rot}. At the same time
the theory is simple in that it does not have local degrees of
freedom. Thus, effectively, one has to deal with only a finite number 
of the degrees of freedom, which makes the usual quantum mechanics
applicable. This gives hope for an explicit construction of the
quantum theory. If constructed, the quantum theory can be expected
to be a non-trivial playground for testing various ideas about 
quantum gravity. Thus, one can expect to give definite
answers to such important conceptual questions as the geometrical
origin of the black hole entropy and information loss in 
black hole evaporation.

A great deal is known about the classical theory, and important advances
have been made in its quantization. We will not be able to give all 
references. Let us only highlight the main developments. The work
\cite{Brown-Hen} of Brown and Henneaux showed that the algebra of asymptotic
symmetries of asymptotically AdS (2+1) gravity is the infinite
dimensional Virasoro algebra of central charge $c=3l/2G$.
The quantum theory should therefore contain the same Virasoro algebra
among its symmetries, and is thus a conformal field theory of this
central charge. Work \cite{Hen-Liouv} used the Chern-Simons (CS) formulation
of the theory to show that asymptotic degrees of freedom can be encoded
in a single scalar field on the boundary of spacetime. The effective
theory that appears on the boundary is Liouville field theory (LFT). 
Subsequent works \cite{Carlip,Banados,Banados-Ortiz,Kaul} also 
used the CS formulation and clarified the question of what are
the degrees of freedom (DOF) in asymptotically AdS (2+1) gravity. Works
\cite{SS,RS} were important in understanding the DOF in the
geometrodynamics formulation. The physical picture that can be extracted
from all these works, and also works on AdS${}_3$/CFT correspondence in
string theory, is that the quantum theory
can be formulated holographically as a CFT living on the boundary of
the spacetime, and this CFT is expected \cite{Hen-Liouv} to be the 
quantum Liouville theory. However, as we explain in section 
\ref{sec:remarks} the actual boundary theory is {\it not}
Liouville theory, although it is closely related to LFT. Most of
the constructions in the present paper are general enough so that
it does not matter what CFT is used. We shall thus refer to this
CFT as Liouville theory, although one should keep in mind the
remarks of section \ref{sec:remarks}.

In the present paper we further develop this ``holographic''
picture. From the technical
point of view, the main novelty will be the usage of quantum LFT on
Riemann surfaces for calculations in the Lorentzian
signature quantum gravity. Thus, the quantum theory is
developed using a suitable analytic continuation procedure.

To motivate our prescription we,
following Maldacena \cite{Malda-Eternal}, consider
Hartle-Hawking (HH) states. A HH state \cite{HH} is a wave-functional
$\Psi[{}^{(2)}\!g]$ that depends on the metric ${}^{(2)}\!g$ on a
spacelike slice $X$. This wave-functional is obtained as the path integral
over Euclidean signature metrics ${}^{(3)}g$ on a manifold $M$ whose
boundary $\partial M=X$:
\begin{eqnarray*}
\Psi[{}^{(2)}\!g] = \int {\cal D}{}^{(3)}\!g\,\, e^{-S_{\rm E}[{}^{(3)}\!g]}.
\end{eqnarray*}
The integral should be taken over metrics ${}^{(3)}\!g$ whose
restriction on the boundary $X$ is ${}^{(2)}\!g$. As we shall explain
in detail below, the spacelike slices $X$ of a typical spacetime are
Riemann surfaces. Thus one should just take ``the
interior'' of $X$ to be $M$. In other words, $M$ is the ``solid'' $X$.
Such $M$ is not unique. Let us, however, at this stage be schematic.
We shall return to the issue of interpretation of different $M$'s below.
It is not hard to evaluate the path integral. As the AdS/CFT 
correspondence suggests, and as was argued in 
\cite{Teich}, the Euclidean path integral over $M$ is equal to
some CFT partition function on $X$. The CFT here is expected \cite{Hen-Liouv}
to be the quantum Liouville theory, or a close relative thereof, see below.
Thus, the HH state is essentially given by the CFT partition function
on the spatial slice $X$. In the bulk of the paper we give details
of this construction and explain what Riemann surfaces arise as
$X$. A special care is necessary in case of a rotating spacetime,
for then there is no natural spatial slice to consider. 
Our construction of HH states thus generalizes and clarifies the proposal of 
\cite{Malda-Eternal}. 

An important new step is to introduce point particles. It
has been proposed in \cite{Part} that point particles should
be described by LFT vertex operators. These are operators of
the form $e^{\eta\phi/b}$, where $\phi$ is the Liouville field,
and $b$ is the Liouville coupling constant.
As is well-known, for $\eta$ real such an operator can be
thought of as making a conical singularity of the deficit angle
$4\pi\eta$ in the surface.
In other words, at least semi-classically, the LFT partition
function on $X$ with such an operator inserted equals to the partition
function on the same $X$ but with a conical singularity at
the point of insertion. Since point particles are essentially
conical singularities is it natural to describe them by these
vertex operators. Generalizing the construction of HH
states to the case when point particles are present, we have
that the states are given by correlation functions on a 
Riemann surface $X$.

One can now interpret HH states as quantum amplitudes. As we discuss
in detail in the bulk of the paper, simplest HH states describe
amplitudes of, for example, black hole creation out of two
point particles, or emission of a particle by a black hole. Taking
the amplitude norm squared one gets the probability for these
processes, which is of direct physical interest. 

We then go onto constructing the S-matrix for point particles.
First we define a natural scalar product between
HH states. The main idea of this product is as follows. A HH state is the
CFT partition function on $X$. The surface $X$ is a Riemann
surface with boundary, one boundary component for every
asymptotic region. There is then a natural inner product between
two HH states for $X_1, X_2$ that have the same number of
asymptotic regions. One just glues two surfaces with boundary
to obtain a closed Riemann surface. The
CFT partition function on this closed surface is the value
of the inner product between the two states. The scalar product
so defined gives transition, or topology changing amplitudes,
see more on this interpretation below.

We use this proposal to define scattering amplitudes
for point particles. These amplitudes are given by the CFT
partition function on a closed Riemann surface with insertion
of a number of vertex operators describing particles. Thus, the
S-matrix for particles is given by a CFT partition function
with insertions. Intermediate states in this
scattering can be interpreted as black holes produced by particles.
The genus of the Riemann surface on which CFT correlator is calculated
is related to the number of asymptotic regions and the
internal topology of the black hole. 

The last step in our construction is to form the full S-matrix
for particle scattering. It is obtained by summing over all
possible ways that an initial configuration of particles can
go to the final one. In particular, one has to sum over
all black hole intermediate states that can be produced by 
particles. The resulting S-matrix is given by the same
prescription that is used in string theory. Namely, it 
is given by the sum over genera integral over the moduli of a 
CFT correlator. The corresponding formula is given in the main 
body of the paper. 

Thus, somewhat surprisingly, we obtain a stringy
S-matrix for point particles. We should emphasize that
this S-matrix is only similar to, but does not coincide with 
the S-matrix of any of the critical string theories.
In particular, what we get is a non-critical CFT with the 
non-zero Brown-Henneaux value of $c$. We must note that,
our CFT being the Liouville theory, the non-critical 
string theory we get is essentially 2d gravity.

The paper is organized as follows. In the next section we
review the classical description of black hole and
point particle spacetimes. An important technical and
conceptual tool is that of an analytic continuation,
which we describe in Section \ref{sec:cont}. We then
define HH states in Section \ref{sec:HH}. Section
\ref{sec:ampl} uses these HH states to obtain probabilities
of simple processes involving point particles. We
discuss the probability of a black hole creation by
point particles and the Hawking emission process.
The S-matrix for point particles is defined in
Section \ref{sec:S-matrix}. We conclude with a 
discussion of open problems and future directions.

\section{Black hole and point particle spacetimes}
\label{sec:class}

In this section we review how the black hole and point particle
spacetimes are constructed. Our main references here are
\cite{Brill,Rot}, see also \cite{Cont} and the companion paper
\cite{BH-Creation}. For more information
on point particles see \cite{Welling} and references therein.

Since there are no local DOF in 3d gravity, all spacetimes look
locally like the maximally symmetric one, that is AdS${}_3$.
Let us thus start by reminding the reader some very basic facts about
the Lorentzian AdS${}_3$, out of which more complicated spaces
will be obtained by identifications of points. The spacetime is
best viewed as the interior of an infinite cylinder. The cylinder
itself is the conformal boundary $\cal I$ of the spacetime. It is
timelike, unlike the null conformal boundary of an asymptotically
flat spacetime. All light rays propagating inside AdS start and
end on $\cal I$. In this picture the constant time slices are
copies of the Poincare (unit) disc. The unit disc is isometric to
the upper half plane $\bf U$; we shall use both models. The
isometry group of the Lorentzian signature AdS${}_3$ is
$\SL(2,\R)\times\SL(2,\R)$. The spacetimes itself can be viewed as
the (universal cover of the) $\SL(2,\R)$ group manifold, and 
the isometry group acts by 
the left and right multiplication. For more details
on AdS${}_3$ see any of the references \cite{Brill,Rot,Cont}.

Let us now turn to the black hole spacetimes. We start with
non-rotating spacetimes. The description of
the non-rotating black holes is greatly facilitated by the fact
that there is a surface $t=0$ of time symmetry. This surface is
preserved by the discrete group $\Gamma$ one uses to identify
points. Thus, $\Gamma$ is actually a subgroup of the group
$\SL(2,\R)\subset\SL(2,\R)\times\SL(2,\R)$ that fixes the $t=0$
plane. This ``diagonal'' $\SL(2,\R)$ consists of transformations
of the form ${\bf x}\to g {\bf x} g^T, g\in\SL(2,\R)$, where we
imply the $\SL(2,\R)$ group manifold model of AdS${}_3$.
Note that this is not the usual diagonal
$\SL(2,\R)$ consisting of transformations ${\bf x}\to g{\bf
x}g^{-1}, g\in\SL(2,\R)$, which fixes the origin of AdS${}_3$.
Transformations fixing the $t=0$ plane act on it by isometries.
Thus, the geometry of the surface $t=0$ is that of the quotient of
the unit disc by the action of $\Gamma\subset\SL(2,\R)$. Such 
2-manifolds are nothing but Riemann surfaces.
Once the geometry of the $t=0$ plane is
understood one just ``evolves'' the identifications in time to
obtain a spacetime, see \cite{Brill}.

\begin{figure}
\centerline{\hbox{\epsfig{figure=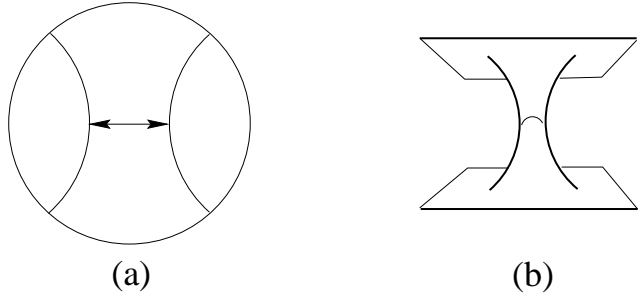,height=1.5in}}}
\caption{BTZ black hole: geometry of the time symmetry surface.}
\label{fig:btz}
\end{figure}

Let us see how this works on examples. Consider first the case of the
non-rotating BTZ BH. In this case the discrete group is generated by a single
hyperbolic element. Its action on the $t=0$ plane can be
understood by finding the fundamental region. In the case
of $\Gamma$ generated by a single element $A$ the fundamental region
is that between two geodesics on $\bf U$ mapped into one another
by $A$, see Fig.~\ref{fig:btz}(a). It is clear
that the quotient space has the topology of the
$S^1\times\R$ wormhole with two asymptotic regions,
each having the topology of $S^1$, see Fig. \ref{fig:btz}(b).
The BTZ angular coordinate runs from one
geodesic to the other. The distance between the two
geodesics measured along their common normal
is precisely the horizon circumference. It can also be determined from
the trace of the generator:
\be\label{trace-bh}
{1\over 2}{\rm Tr}\, A = \cosh{\pi r_+}.
\ee
The black hole mass is then 
\be
M/\pi = 1+ r_+^2.
\ee
Here we work in the units $8\pi G=l=1$, where $l=1/\sqrt{-\Lambda}$. 
We take the empty AdS${}_3$ to have zero mass. The ``zero mass'',
or, more correctly, zero size black hole then corresponds to 
$M=\pi$. Having understood
the time symmetry surface geometry, one can obtain
the spacetime geometry by ``evolving'' in time the $t=0$ slice,
see \cite{Brill}. One finds that the resulting spacetime is
indeed a black hole, in the sense that there is a region
causally disconnected from the asymptotic infinity.

\begin{figure}
\centerline{\hbox{\epsfig{figure=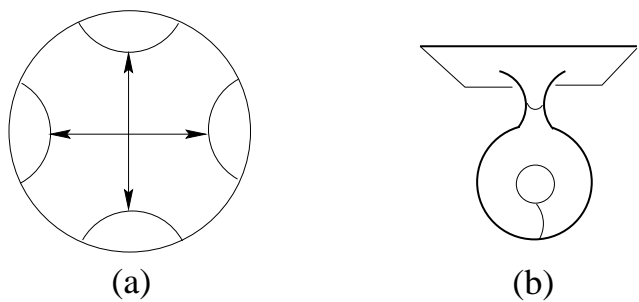,height=1.5in}}}
\bigskip
\caption{Time symmetry surface geometry of the single asymptotic region
black hole with a torus wormhole inside the horizon}
\label{fig:wormhole}
\end{figure}

Let us consider more complicated initial slice geometries.
We now consider the group $\Gamma$ to be generated by two
hyperbolic elements. For example, let the fundamental region be
the part of the unit disc between four geodesics, as in
Fig. \ref{fig:wormhole}(a). Let us identify these geodesics
cross-wise. It is straightforward to show that the resulting
geometry has only one asymptotic region, consisting of all
four parts of the infinity of the fundamental region. With
little more effort one can convince oneself that the resulting
geometry is one asymptotic region ``glued'' to a torus,
see Fig. \ref{fig:wormhole}(b). The spacetime
obtained by evolving this geometry is 
a single asymptotic region black hole, but the topology
{\it inside} the event horizon is now that of a torus.
See \cite{Brill} for more details on this spacetime.

\begin{figure}
\centerline{\hbox{\epsfig{figure=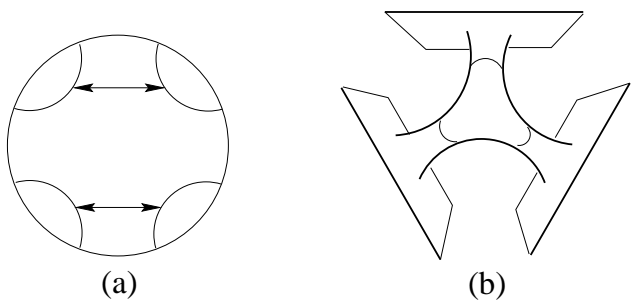,height=1.5in}}}
\caption{Symmetry surface geometry of the three asymptotic region black hole}
\label{fig:3bh}
\end{figure}

A group generated by two elements can also be used to obtain
a three asymptotic region black hole \cite{Brill}. The fundamental
region on the $t=0$ plane is again the region bounded by four
geodesics. They are, however, now identified side-wise,
see Fig. \ref{fig:3bh}(a). One can clearly see that the
initial slice geometry has three asymptotic regions, with
the black hole separated from the asymptotic regions by three
horizons, see in Fig. \ref{fig:3bh}(b). Evolving this, one gets a spacetime
with three asymptotic regions and corresponding event horizons.
See \cite{Brill} for more details.

Taking the group $\Gamma$ to be more complicated
one constructs a large class of spacetimes.
In particular, one can have a single asymptotic region
black hole with an arbitrary Riemann surface inside
the horizon. More generally, one can have a black hole
with any number of asymptotic regions, and with any number of
handles hidden behind the horizon(s).

To get point particle spacetimes one has to take some of the generators
to be elliptic. Let us again consider the case of a non-rotating
spacetime with a time symmetry plane $t=0$. This describes point
particles that are at rest at $t=0$. The spacetime describing
a single point particle is obtained by taking the group $\Gamma\in\SL(2,\R)$
generated by a single elliptic element $A$. The elliptic $A$ has
a fixed point inside the unit disc. After identifications are carried
out this point becomes a conical singularity. The deficit angle
is determined by the holonomy of ${\rm Tr}\,A$. We have:
\be\label{trace-part}
{1\over 2}{\rm Tr}\,A = \cos{\pi\alpha},
\ee
and the deficit angle is $4\pi\eta = 2\pi(1-\alpha)$. This describes
a point particle of mass
\be
M/\pi = 1-\alpha^2.
\ee
Particle's mass is then
bounded from above by $M=\pi$. A particle of this maximal mass 
corresponds to a parabolic generator and can
be viewed as the ``zero mass'' black hole.  The geometry of
surface ${\bf U}/\Gamma$ is that of a cone over a circle.
The tip of the cone is a singularity. It should be thought of as the
point where the worldline of the particle intersects the $t=0$
plane. 

Spacetimes that contain both black holes and point particles are
obtained by using a group $\Gamma$ generated by several elements of 
different type. One has to use Klein combination theorems, see e.g.
\cite{Krush}, to select individual generators in such a way that
the resulting time symmetry surface has the desired topology. 
In particular, one can have any of the black holes
discussed above with one (or several) point particles crossing the
time symmetry surface. The crossings can occur both behind the horizon
and in the asymptotic regions. For example, one can have a BTZ black hole
with a point particle located outside of the horizon, 
in one of the asymptotic regions. This configuration will be used in our
description of the Hawking emission process.

Let us now turn to the rotating case, that is to the case with
no plane of time symmetry in the spacetime. To describe 
and classify these spacetimes to the extent we 
understand non-rotating ones it turns out to be convenient
to perform an analytic continuation \cite{Cont}. We describe
this in the next section.

\section{Analytic continuation}
\label{sec:cont}

The material presented in this section is from \cite{Cont}.

The idea is to analytically continue the discrete group 
$\Gamma\subset\SL(2,\R)\times\SL(2,\R)$ to some discrete group
$\Sigma$ of $\SL(2,\C)$, which is the isometry group of
the Euclidean AdS${}_3$, same as the hyperbolic space $\H$. 
One then uses $\Sigma$ to obtain a quotient $M=\H/\Sigma$.
Hyperbolic 3-manifolds such as $M$ are well understood.
One uses this knowledge to understand and classify the
rotating spacetimes. More precisely, it is best to concentrate on the 
conformal boundary $\tilde{X}$ of the resulting 3-manifold $M$. This is
a Riemann surface, and its topology and the moduli are
in one-to-one correspondence with the geometry of the
corresponding black hole spacetime. 

\begin{figure}
\centerline{\hbox{\epsfig{figure=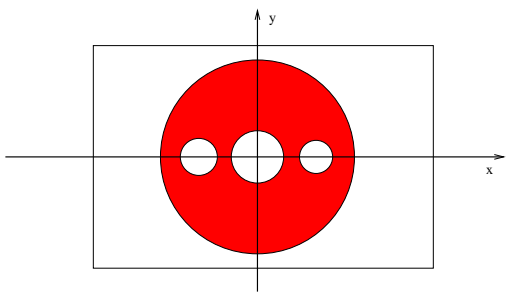,height=1.5in}}}
\caption{The fundamental domain for the Schottky uniformization
of a genus 2 surface.}
\label{fig:Schottky}
\end{figure}

Let us first cast the non-rotating case in this 
language. The group $\Gamma$ is then a
subgroup of $\SL(2,\R)$. The analytic continuation
prescription in this case is to take the {\it same}
group $\Gamma$ but think of it as a subgroup of
$\SL(2,C)$. There is then a simple relation between
the geometry of the time symmetry plane and the
conformal boundary $\tilde{X}$ of $M=\H/\Gamma$. It turns
out that $\tilde{X}$ is simply the Schottky double of the
geometry of the time symmetry plane $X$. Let us remind the
reader that the Schottky double exist for any Riemann
surface. For a compact surface $X$, the double
$\tilde{X}$ is simply two disconnected copies of
$X$, with all moduli replaced by their complex conjugates
in the second copy. For a Riemann surface with 
boundary, as is the case for the time symmetry surfaces
of our black holes (boundaries are asymptotic regions),
the double is obtained by taking two copies of $X$,
taking complex conjugates of all the moduli in the second copy,
and gluing them along the boundary to obtain a closed connected
Riemann surface $\tilde{X}$. The Schottky double is obtained
exactly by uniformizing the surface $X$ by a Fuchsian group
$\Gamma$, and then acting by $\Gamma$ on the whole complex 
plane instead of the upper half-plane. The double
$\tilde{X}$ is simply the quotient ${\cal C}/\Gamma$,
where $\cal C$ is the complement of the set of fixed points
of action of $\Gamma$ on $\C$. As an example, let us give a
picture of the fundamental domain for the action on $\C$ of
the group $\Gamma$ generated by two hyperbolic generators.
The upper half of the Fig.~\ref{fig:Schottky} is simply
the fundamental domain on the unit disc, Fig.~\ref{fig:wormhole}(a)
or Fig.~\ref{fig:3bh}(a), mapped into the upper half-plane.
Considering the action on the full complex plane, and taking
the quotient, one gets a closed Riemann surface, in this case
of genus two --the Schottky double of $X$.

Having explained why the boundary of the Euclidean space is
the Schottky double of the initial slice geometry, let us use this
to give a simple relation between the number of asymptotic
regions $K$, the number of handles $G$ behind the horizon, and the genus $g$
of the Euclidean boundary $\tilde{X}$. As is not hard to see:
\begin{equation}
g = 2G + K - 1.
\end{equation}
We see that several different time symmetry surface geometries
can correspond to the same genus of the double $\tilde{X}$.
This is because Riemann surface $\tilde{X}$ can be cut into
two ``equal'' pieces in several different ways. This is somewhat
reminiscent of string theory, where a single Riemann surface
corresponds to many different field theory Feynman graphs, depending
on how one decomposes the surface into pants. We shall see more
parallels with string theory in what follows. 

Let us note that curves along
which one cuts $\tilde{X}$ to obtain two copies of $X$ correspond to 
asymptotic regions on $X$. In particular, the holonomy along
such a curve $\alpha$ (that is the trace of an element of $\Gamma$ whose
axis projects on $\alpha$) determines, see
(\ref{trace-bh}) the size (and the mass) of the
horizon separating this asymptotic region from the rest of spacetime.
Thus, to distinguish different spacetimes that give rise
to the same Euclidean boundary $\tilde{X}$ one should mark
$\tilde{X}$ with a set of homotopy non-trivial non-intersecting
curves that divide $\tilde{X}$ into two pieces of
equal topology and correspond to asymptotic regions.

Let us discuss what kind of 3-manifold $M$ one gets
as $\H/\Gamma$. Obtained by identification of points in $\H$,
$M$ is a constant negative curvature 3-manifold. Its conformal
boundary is precisely the surface $\tilde{X}$. Let us note, however, 
that given a Riemann surface $\tilde{X}$ there are many different
3-manifolds $M$ whose boundary is $\tilde{X}$. To
get a specific manifold one has to select on $\tilde{X}$ a
basis of $\pi_1(\tilde{X})$ consisting of curves 
$\alpha_i,\beta_i$. Such surface $\tilde{X}$ is called
marked. There is then a single 3-manifold $M$ whose
boundary is $\tilde{X}$ and such that curves $\alpha_i$
are all contractible inside $M$. This manifold is obtained
by the Schottky uniformization of $\tilde{X}$. Namely, given
a marked Riemann surface $\tilde{X}$ there is a unique
Schottky group $\Sigma$, that is a group freely generated
by $g$ elements $L_1,\ldots,L_g$, such that $\tilde{X}$
is given by the quotient of the complex plane by $\Sigma$
and all generators $L_i$ project to the curves $\beta_i$.
One can then use this Schottky group to obtain a hyperbolic
3-manifold $M=\H/\Sigma$. The curves $\beta_i$ are non-contractible
inside $M$, whereas $\alpha_i$ are contractible.

In our case the manifold $M$ is a specific one, given by
$M=\H/\Gamma$. In practice $\Gamma$ comes with a set of
generators chosen, in other words it is marked, and the axes of these
generators project to curves on $\tilde{X}$ that are non-contractible
inside $M$. In particular, the curves on $\tilde{X}$ that correspond
to asymptotic regions, that is divide $\tilde{X}$ into two parts
$X$, are all either projections of the axes of generators of
$\Gamma$, or are obtained as commutators of the generators.
In any case they are non-contractible inside $M$ due to the
presence of horizons. Thus, $M$ can be pictured as the
``exterior'' of $\tilde{X}$ in $S^3$, not as the interior.

Let us now generalize this picture to rotating
spacetimes. The main idea is to take the group
$\Gamma$ of a non-rotating spacetime and deform
it into a subgroup of $\SL(2,\C)$, this deformation
taking into account the rotation. As we showed in
\cite{Cont}, it is natural to consider the 
Fenchel-Nielsen deformation. The result of the
deformation is a quasi-Fuchsian group $\Gamma^\tau$.
The Euclidean boundary $\tilde{X}$ is then
the quotient of the complex plane with respect to
the transformations from $\Gamma^\tau$. Geometrically,
turning on the rotating in one of the asymptotic regions
amounts to performing the Fenchel-Nielsen
twist on the Riemann surface $\tilde{X}$ along the
geodesic $\alpha$ that corresponds to this asymptotic region.
The angular velocity of the asymptotic region is
determined by the (imaginary part of the) trace of an element of $\Gamma^\tau$
whose axes projects on $\alpha$. This brief description
will be sufficient for our purposes. See \cite{Cont} for
more detail. 

Let us summarize. A non-rotating black hole is described in the
Euclidean signature by the Schottky double $\tilde{X}$
of the time symmetry plane $X$. The closed Riemann surface
$\tilde{X}$ is the conformal boundary of a hyperbolic 3-manifold
$M$ --analytic continuation of the black hole spacetime.
A set of geodesics on $\tilde{X}$ that separates it into
two parts is in one-to-one correspondence with asymptotic
regions of $X$. One turns on the rotation by performing
the Fenchel-Nielsen twist along these geodesics. The holonomy 
along these geodesics determine the size and rotation of the 
corresponding horizons. This analytic continuation construction
gives an explicit description and classification of the rotating 
spacetimes. Whereas the non-rotating spacetimes are classified by 
their time symmetry plane $X$ geometry, rotating black holes are 
classified by the geometry of the double $\tilde{X}$.

Although we have only described the case of hyperbolic generators,
that is no point particles, the above discussion 
generalizes to the case when particles are present. When the
spacetime is non-rotating, and the particles are in rest
on the time symmetry surface, one has a group $\Gamma\subset\SL(2,\R)$
that contains elliptic (and/or parabolic) elements. One can
again consider the action of $\Gamma$ on the whole complex
plane and obtain the double $\tilde{X}$. It is 
a closed surface with conical singularities, whose
number is twice the number of singularities on the
time symmetry surface. The position of singularities
on $\tilde{X}$ is obtained from the position on $X$
by taking the mirror image with respect to the real axes. The case of
rotating spacetimes, or case of particles with momentum
is more complicated. One can still form a double $\tilde{X}$
that has the same topology (and the same number of insertions) 
as in the non-rotating case. However, the position of
insertions on $\tilde{X}$ is more complicated to determine.
It depends both on the rotation of the spacetime and on the momentum of
particles. In the present paper, being rather schematic,
we will not need a precise relation. It is given in the 
companion paper \cite{BH-Creation}.

\section{Hartle-Hawking states}
\label{sec:HH}

Having reviewed how the spacetimes in question are described,
we are ready to consider quantum theory. As we have explained
in the introduction, the idea is to consider HH states. In
the context of negative cosmological constant gravity in (2+1) 
dimensions such states were considered in \cite{Malda-Eternal}.
The author considered only simplest of such states. In this
section we generalize his considerations, in particular to spacetimes
containing point particles and to the rotating case. 
The next section considers some simple physical processes that
can be analyzed with the help of the amplitudes we define here.
Section \ref{sec:S-matrix} defines an inner product for 
HH states.

Let us consider the non-rotating case first. In this case there
is a plane $t=0$ of time symmetry. It is natural to
define HH states using the plane $X$. Thus, a state $\Psi$ will
be a quantum state of our spacetime at time $t=0$. It is defined
as the path integral over metrics of Euclidean signature
on a 3-manifold $M$ whose boundary is $X$. There are different
possible choices of such $M$. 
Let us for now be schematic, leaving a discussion on the choice of
$M$ till later. At this stage one should think of $M$ as
being the ``interior'' of the Riemann surface $X$.
We will now utilize the result of \cite{Teich} which states
that the Euclidean path integral over a 3-manifold $M$ whose
boundary is $X$ is equal to the CFT partition function on $X$:
\be\label{3D-cft-relation}
\int_M {\cal D}{}^{(3)}\!g \,\, e^{-S_{\rm E}[{}^{(3)}\!g]} = 
Z_{\rm CFT}[X].
\ee
This result is, of course, in the general spirit of AdS${}_3$/CFT
correspondence, specialized to the case when no fields except for
the metric is present. The relation (\ref{3D-cft-relation}) 
was obtained in \cite{Teich} by using the CS formulation
of $\Lambda<0$ Euclidean 3D gravity. It was shown that
the gravity partition function holomorphically factorizes,
and thus is a partition function of some CFT. There are 
reasons to believe \cite{Hen-Liouv} that this CFT is
related to the quantum Liouville theory. For
now we shall refer to this CFT as simply Liouville field
theory (LFT), but we note that the relevant CFT is rather
a certain close relative of LFT that incorporates the
point particle states, see section \ref{sec:remarks}
for a discussion on this point. 

We must note that the result
(\ref{3D-cft-relation}) was obtained in \cite{Teich} for the
case of a closed Riemann surface $X$, while spatial
slices that are relevant for our purposes have boundaries
(asymptotic regions). Some extra thought is required to
establish an analog of (\ref{3D-cft-relation}) for surfaces
with boundaries. Let us for the moment analyze the case of 
a closed $X$. Such spacetimes are simple cosmological models,
and were considered in this context in, e.g., \cite{Gary}.
Thus, for the case of a non-rotating spacetime with a closed 
time symmetry plane $X$ the HH state is given by
the Liouville partition function on $X$.
Note that $Z_{\rm CFT}[X]$ only depends non-trivially
on the conformal structure of $X$, that is, it is a 
function on the Teichmuller space $T_X$ of $X$, not
on the space of metrics ${}^{(2)}\!g$ on $X$. 

One can interpret the fact that $\Psi[{}^{(2)}\!g]$ only depends on
the conformal structure, not on the full metric ${}^{(2)}\!g$ 
as follows. Recall that in
the canonical approach to quantum gravity a quantum state $\Psi$ 
should satisfy the Wheeler-DeWitt equation. In other
words, a physical state must be annihilated by the
constraints. A way to get such states is to first find
the reduced phase space at the classical level, and
then consider functions on the reduced configuration
space. It is known that the reduced phase space of 2+1
gravity is the cotangent bundle over the Teichmuller space 
of the spatial slice.
In the context of zero cosmological constant this is
a well-known result due to Moncrief \cite{Moncrief}.
In our negative cosmological case one can understand
this result as follows. Let us use the CS formulation.
Then $\Lambda<0$ Lorentzian signature gravity is
$\SL(2,\R)\times\SL(2,\R)$ CS theory. The reduced
phase space of CS theory on a manifold of topology
$X\times\R$ is the space of homomorphisms 
$\pi_1(X)\to\SL(2,\R)\times\SL(2,\R)$. This space
is parametrized by holonomies of the two connections
along the loops generating $\pi_1(X)$. One can
introduce complex coordinates on this
space by performing the analytic continuation of
$\SL(2,\R)\times\SL(2,\R)$ to $\SL(2,\C)$,
in the way we explained in the previous section,
and that is described in more detail in \cite{Cont}.
Under this continuation the CS symplectic structure goes
to the CS symplectic structure on the 
the space of homomorphisms $\pi_1(X)\to\SL(2,\C)$.
This later space is known to be the same as the
space of projective structures on $X$, which is
the same as the cotangent bundle the Teichmuller
space $T_{\rm X}$, see e.g. \cite{Kawai}.
The reduced phase space being $T^* T_{\rm X}$
explains why it is natural for the quantum states
$\Psi$ to be functions on $T_{\rm X}$.

Having said this, there are two cautionary remarks in order.
First, because the partition function is modular invariant it
is a function on the moduli (Riemann) space ${\cal M}_X$
rather than on $T_X$. Second, the CFT partition function
is not really a function, but rather a section of a line
bundle over ${\cal M}_X$. Indeed, due to the presence of
the conformal anomaly $Z_{\rm CFT}$ depends not
only on the moduli, but also on a representative of the metric
${}^{(2)}\!g$ on $X$. Under a change of this representative
by a conformal rescaling ${}^{(2)}g\to e^\varphi{}^{(2)}\!g$ 
the partition function gets multiplied by the
exponent of the classical Liouville action for $\varphi$. 
Thus, to make $Z_{\rm CFT}$ a function on ${\cal M}_X$ we
have to select a representative ${}^{(2)}\!g$ for
every point in ${\cal M}_X$. A natural representative is the
canonical (Poincare) metric of constant negative
curvature that exists (and is unique) on every $g>1$
Riemann surface. When we talk about the partition function
as a function on ${\cal M}_X$ we will always imply this choice. 

It seems to be the right time to discuss the issue of the choice
of $M$ that needs to be made to compute the path
integral in (\ref{3D-cft-relation}). For the case of a closed
$X$ that we are presently discussing, there are different
possible choices that can be distinguished by specifying
which set of loops on $X$ is contractible inside $M$. 
One needs to specify $g$ such homotopy non-trivial loops
on $X$, where $g$ is the genus of $X$. Any two such sets of loops
are related by a modular transformation on $X$. On the other hand, the
right hand side of (\ref{3D-cft-relation}), that is, the
CFT partition function, is modular invariant. This means
that the left hand side should be interpreted as containing 
not only the path integral over metrics on $M$ of a fixed topology, 
but also a sum over different 3-manifolds $M$ that have the same conformal
structure on $X=\partial M$. Such a sum over topologies 
is by now standard in the setting of AdS/CFT correspondence,
see \cite{Witten-AdS} for the first discussion of this issue.
Thus, for a closed $X$, a choice of $M$ in (\ref{3D-cft-relation})
is not important, one should sum over all possibilities.

Let us now return to the case relevant for our purposes, that is,
when Riemann surface $X$ has a boundary. There is one circular boundary 
component for every asymptotic region. Motivated by
the result (\ref{3D-cft-relation}) valid for a closed surface $X$,
we define an open $X$ HH state to be the CFT partition function on $X$
as well. From the discussion that follows it will become clear 
why this definition is natural. We shall see, however,
that there are some important subtle differences from the closed
case.

To evaluate the  
CFT partition function on $X$ with boundary one has to specify boundary
conditions. In the case of rational CFT there
is a special subset of boundary conditions, the
one considered in the original work by Cardy 
\cite{Cardy}. These Cardy boundary conditions
are in one-to-one correspondence with 
primaries of the CFT. For a non-compact CFT,
as is Liouville theory, one has 
similar Cardy boundary states,
see \cite{Zamolo-B,Teschner-B1,Teschner-B2} for
a recent discussion of boundary LFT. The
LFT boundary states are labelled by a single
real number, for which we shall employ
lower case Latin letters. Specifying 
this boundary label for every asymptotic region
one gets Riemann surface with labels. The HH
state is defined as the Liouville partition function on such $X$.
Schematically,
\be\label{state-1}
\Psi_{\rm HH}[a,b,c\,] = Z_{\rm CFT}\left[\,\,
\lower0.28in\hbox{\epsfig{figure=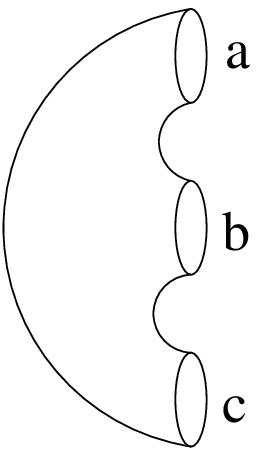,height=0.6in}}
\,\,\right],
\ee
where we have used the geometry of the three asymptotic
region black hole for illustration. Boundaries of $X$
correspond to asymptotic regions. One should
think of the number labelling each boundary as in a certain
sense dual to the size of the corresponding horizon, see
more on this below.

One can also consider the time symmetry surface geometries
containing point particles. The HH states are still given
by the CFT partition function on the corresponding Riemann
surface. The point particles are incorporated by
inserting LFT vertex operators. For example, 
the HH state of the three asymptotic
region black hole with three point particles on the time
symmetry slice is given by:
\be\label{state-1-part}
\Psi_{\rm HH}[\xi_1,\xi_2,\xi_3;a,b,c\,] = Z_{\rm CFT}\left[\,\,
\lower0.28in\hbox{\epsfig{figure=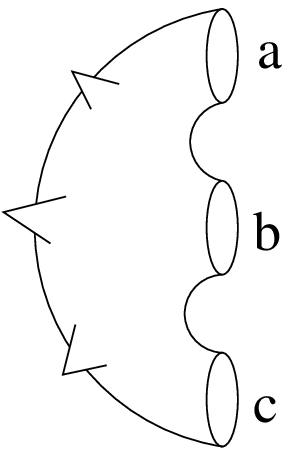,height=0.6in}}
\,\,\right].
\ee
The right hand side can either be interpreted as
the LFT partition function on a 3-holed sphere with
three vertex operators inserted, or as the
partition function of the 3-holed sphere with
three conical singularities. The parameters $\xi_i$
are those of the LFT vertex operators $e^{\xi\phi}$.
In the semi-classical limit $b\to 0$, where $b$ is the
LFT coupling constant, the relevant parameter is
$\eta=\xi b$. The deficit angle
created is equal to $4\pi\eta=2\pi(1-\alpha)$, where
the parameter $\alpha$ is the one introduced for point
particles in (\ref{trace-part}).

Actually, one can similarly
think of the holes as being made by an insertion into the
sphere of non-local LFT vertex operators.
While vertex operators creating conical singularities
are given by $e^{\xi\phi}$, with $\xi$ real,
non-local vertex operators that create holes correspond to
$\xi=Q/2+iP$, where $Q=b+1/b$, and $P$ is proportional to the size of the
hole it makes. Thus, one can obtain the boundary states
(\ref{state-1}) as a linear combination of the
correlation functions with non-local operators inserted. 
This is done by introducing the Ishibashi states.
Schematically,
\be
\langle a | = \int_0^\infty {dP\over\pi} U(P) \langle P |,
\ee
where $\langle P|$ is the Ishibashi state (sum over
the descendants) constructed from
the primary state created by $e^{\xi\phi}$ with $\xi=Q/2+iP$,
and $U(P)$ is the one point function in the boundary LFT,
see \cite{Zamolo-B}. The boundary
condition $a$ is thus dual (in the above sense) to the size
$P$ of the hole.

Let us now introduce another, equivalent definition of the
open $X$ HH states. It is one of the defining properties
of the boundary CFT, see, e.g., \cite{Schw}, that the 
full CFT partition function
on a surface with boundary equals to the chiral CFT
partition function on the Schottky double:
\be\label{chiral-part}
Z_{\rm CFT}[X] = Z_{\rm chiral}[\tilde{X}].
\ee
The partition function on the right hand side is evaluated
in the presence of holonomies. There is one holonomy
for every boundary component. Each holonomy is labelled
by the corresponding boundary condition.
As we have seen in the previous section, the Schottky
double $\tilde{X}$ of the time symmetry plane 
is the boundary of the Euclidean 3-manifold that
is the analytic continuation of the black hole spacetime.
Thus, our HH states also have the interpretation of
the chiral CFT partition function on the Euclidean
boundary. Schematically,
\be\label{def-chiral}
\Psi_{\rm HH}[a,b,c]=Z_{\rm chiral}\left[\,\,
\lower0.28in\hbox{\epsfig{figure=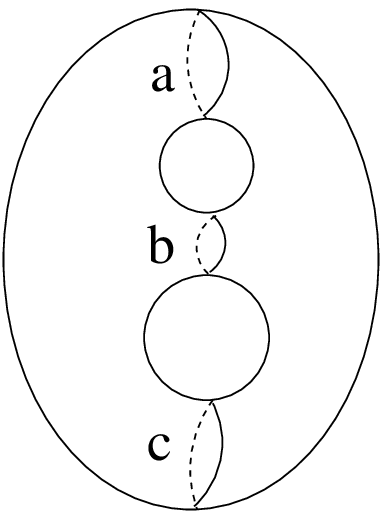,height=0.6in}}
\,\,
\right].
\ee
As is shown in \cite{Teich}, this chiral
CFT is nothing else but the $\SL(2,\C)$
CS theory for one of the two complex conjugate connections
that describe gravity in the Euclidean AdS. The chiral
CFT partition function, or, in other words, a conformal
block, is a particular quantum state of the
CS theory. Thus, the chiral definition (\ref{def-chiral}) of HH 
states makes it clear that they are indeed quantum 
states in that they can be obtained as the path integral
of exponential of $i$ times the $\SL(2,\C)$ CS action.

The definition (\ref{def-chiral}) HH states as chiral CFT partition
function on the double $\tilde{X}$ allows us to be more 
precise on the issue which manifold $M$ to take when
evaluating the Euclidean 3d gravity path integral. As we have
discussed above, for a closed surface $X$ the result is the full 
CFT partition function on $X$, which is modular invariant. 
One sums over all possible $M$'s in the path integral to produce a modular
invariant answer. For an open surface $X$ the situation
is more complicated. We recall that the chiral CFT partition function,
in other words a conformal block, or a state of CS theory,
has complicated transformation properties with respect
to modular transformations. It can be thought of as 
a section of a very non-trivial holomorphic fiber bundle over the moduli
space ${\cal M}_{\tilde{X}}$. It is best, however, to think of
it as a section of a (trivial) holomorphic 
fiber bundle over the Teichmuller space
$T_{\tilde{X}}$. Note that it is important where the holonomies are located
on $\tilde{X}$. Choosing a different set of holonomies for
boundary conditions, or a different labelling of them
changes the HH state (\ref{def-chiral}). Thus, HH states
depend explicitly on a marking of $\tilde{X}$ with a 
set of curves that correspond to asymptotic regions. 
On the other hand (\ref{state-1}) is invariant under
the modular transformations that leave the boundary of 
$X$ (that is, asymptotic regions) invariant.

Having defined $\Psi_{\rm HH}$ as
the chiral CFT partition function on the double $\tilde{X}$,
we have $\Psi_{\rm HH}$ equal to the $\SL(2,\C)$
CS theory partition function on a specific 3D manifold
$\tilde{M}$, whose boundary is $\tilde{X}$. The
manifold $\tilde{M}$ is just the analytic continuation
of the black hole spacetime, that is the space $\H/\Gamma$.
Since the Fuchsian group $\Gamma$ is marked, this is a specific
3D space, with generators of $\Gamma$ corresponding to
asymptotic regions non-contractible inside $\tilde{M}$. 

Another useful interpretation of HH states that results from
(\ref{def-chiral}) is as follows.
As is shown in \cite{Teich}, the phase space of each of the two
$\SL(2,\C)$ CS theories is the holomorphic (or anti-holomorphic)
part of the full phase space, which is the cotangent bundle
over the Teichmuller space of $\tilde{X}$. This 
holomorphic part of $T^* T_{\tilde{X}}$ is itself naturally isomorphic
to $T_{\tilde{X}}$. Thus, our HH states can also
be thought as particular states in the Hilbert space $\cal H$
obtained by quantizing the Teichmuller space of $\tilde{X}$. 
In other words, $\Psi_{\rm HH}$ are particular holomorphic
functions of the moduli of $\tilde{X}$. Different choice
of boundary conditions (or marking of $\tilde{X}$) 
give different states in the same
Hilbert space $\cal H$. Thus, HH states
of the Lorentzian signature theory are particular states in
the Hilbert space obtained by quantizing 
the Teichmuller space of the Euclidean boundary $\tilde{X}$.
This makes the theory of quantum Teichmuller spaces developed in
\cite{Fock,Kashaev} directly relevant to quantum gravity in
(2+1) dimensions.

This last interpretation allows for an immediate generalization
of HH states to the rotating case. On
the first sight this seems to be impossible, for there
is no time symmetry plane in this case. However, as we
saw above, the non-rotating HH states can also be interpreted
as states in the Hilbert space ${\cal H}$ of the quantum Teichmuller 
space of the Schottky double $\tilde{X}$. Recall now that
the rotation can be incorporated as the Fenchel-Nielsen twist
on $\tilde{X}$. One gets a new Riemann surface $\tilde{X}^\tau$ of
the same genus, but with different values of the moduli.
Thus, the same Hilbert space $\cal H$ obtained by quantizing
the Teichmuller space $T_{\tilde{X}}$ also contains states
that correspond to rotating spacetimes. They are just different
holomorphic functions of the moduli of $\tilde{X}$. One should be able to
obtain these states by evaluating the chiral CFT partition
function (\ref{chiral-part}) in the presence of more 
general holonomies. It can be expected that the general
holonomy (and thus a boundary condition) in Liouville
CFT is characterized by a single complex number.
Such more general boundary condition is dual to a more
general vertex operator $e^{\alpha\phi}$ with
$\alpha$ not necessarily of the form $\alpha=Q/2+iP$.
The conformal dimension for states created by such
operators is complex. The imaginary part of this
conformal dimension is the angular momentum of the
corresponding asymptotic region. 
We will not need details on the rotating states
in the discussion that follows. It suffices
to know that they exist.

\section{Quantum amplitudes for processes involving point particles}
\label{sec:ampl}

Having understood how the HH states are constructed, we can
use them to study simple physical processes. There are
two main processes of interest: (i) emission (absorption) of a 
point particle by the BTZ black hole, and (ii) creation
of a black hole by two point particles. In this section we
explain how the quantum probabilities for these processes
can be obtained. We will be rather schematic and give no
detailed calculations. Details on the BH creation process
are given in the companion paper \cite{BH-Creation}.

\subsection{Emission of a point particle by the BTZ BH}
\label{sec:emission}

Black holes emit particles. Having point particles in our
theory, we expect that there is an emission process in
which a point particle gets created in the vicinity
of the horizon and propagates away from the hole. Only massless 
particles can reach infinity in AdS. Thus, there is a strict version of
the Hawking process only for lightlike particles. 
Massive particles are also expected to be created, but they
will eventually fall back into the black hole. There is a
question how to describe these processes, for there is a conceptual 
difference from the original Hawking
\cite{Hawk} setup for black hole evaporation. Indeed, there are
no extra fields present to carry away radiation. Thus, one
can neither quantize fields in the black hole background,
which is the original Hawking derivation \cite{Hawk}, nor
consider the hole as a quantum mechanical system in external
fields, see, e.g., Bekenstein \cite{Bek}. A derivation that
is applicable to our case is that of \cite{Wilzek},
which treats the emission process as quantum tunneling.
In this approach the analysis is performed in the geometric optics 
approximation, in which radiation travels along (null) geodesics.
This is directly applicable to our case with quanta traveling
along geodesics being point particles. For the AdS black holes
the analysis was performed in \cite{AdS-Tunn}, see also 
\cite{Italian} for a derivation for the case of BTZ BH.
These results tell us that our black holes are indeed 
expected to radiate point particles.

To describe this emission qualitatively we need to find
a HH wave function whose amplitude squared will give the
probability for the process. According to our prescription,
a HH state is the LFT partition function on the relevant
spatial slice. For the case of emission, this slice has 
the geometry of the infinite throat, with a point particle 
in one of the asymptotic regions. The relevant HH state
is thus the LFT partition function on a two-holed sphere
with a single vertex operator inserted. Each of the
two boundaries corresponding to the asymptotic regions
must be assigned a boundary condition. The two
boundary conditions need not be the same. Thus,
we get a state:
\be\label{bh-emission-amplitude}
\Psi_{\rm HH}[a,b;\eta]=Z_{\rm CFT}\left[\,\,
\lower0.12in\hbox{\epsfig{figure=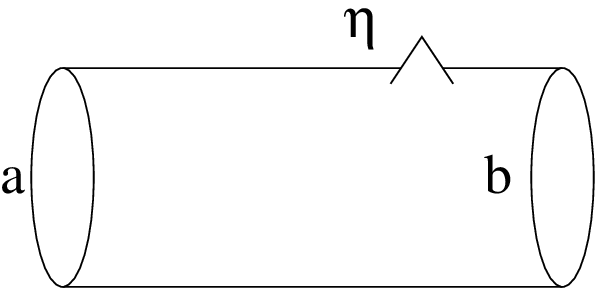,height=0.35in}}
\,\,\right].
\ee
The parameter $\eta$ is determined by the rest mass of the particle
emitted. The state depends, in addition to the two
boundary conditions $a,b$ and the parameter $\eta$, on the
modulus of the cylinder, and on the location of the point of insertion
of the vertex operator. The position of the point of insertion
is determined by the momentum of the particle. In this paper
we won't need a precise relation. 

One can obtain the full probability of emission of a particle
of given mass and momentum by taking the absolute value
square of the amplitudes (\ref{bh-emission-amplitude})
and summing over the boundary conditions. The sum over
boundary conditions glues two copies of the cylinder 
together to produce a torus. The full emission probability
is then given by the CFT partition function on the torus
with two insertions:
\be\label{em-prob}
{\cal P} = \sum_{a,b} Z_{\rm CFT}\left[\,\,
\lower0.12in\hbox{\epsfig{figure=emission.eps,height=0.35in}}
\,\,\right] 
\overline{Z_{\rm CFT}}\left[\,\,
\lower0.12in\hbox{\epsfig{figure=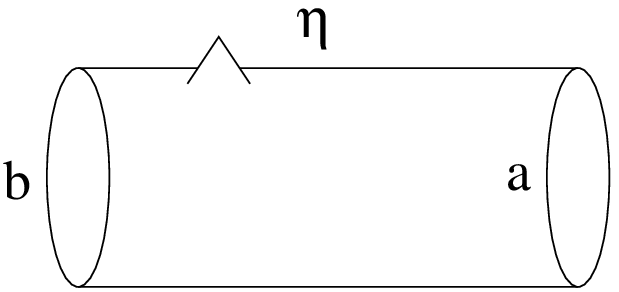,height=0.35in}}
\,\,\right]  = 
Z_{\rm CFT}\left[\,\,
\lower0.11in\hbox{\epsfig{figure=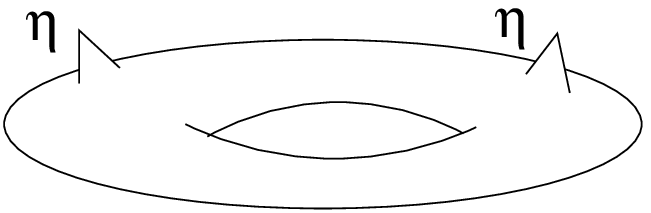,height=0.3in}}
\,\,\right].
\ee
The probability $\cal P$ is a function of the modulus of
the torus, which codes the parameters of the black hole,
of the mass parameter $\eta$, and of the relative position
of the two insertion points, which codes particle's
momentum. 

To see that this is indeed the ``right'' answer for the
emission/absorption probability we recall some facts about
the absorption of a scalar field by the BTZ black hole. The point
is that in the limit of large frequency, this absorption
probability is given by essentially the same two-point
function on the torus. In this limit the geometric optics
approximation is valid, which means that absorption for
waves is the same as absorption for particles. The fact that 
the absorption probability is given by the two-point function 
means that our answer (\ref{em-prob}) is essentially correct.

Absorption probability for scalar (and other) fields 
propagating in the background of the BTZ BH can be found exactly.
This is usually discussed for massless field,
when the field can reach the infinity, see for results
on the minimally coupled massless scalar \cite{Ivo}. For massive particles
the absorption cross-section can be read off from the
AdS/CFT prediction for the emission rate, see \cite{Jap}.
Importantly, in all the cases the absorption probability
can be expressed as a certain integral of the thermal two point function 
in the CFT.  More precisely, as was demonstrated by 
\cite{Malda-S,Gubser} the absorption probability for
a wave with momentum along the boundary equal to 
${\bf p}=(\omega,p)$ is proportional to  
\be\label{abs-prob-int}
\int d^2{\bf x}\, e^{i{\bf p}\cdot {\bf x}} {\cal G}(t-i\epsilon, x),
\ee
where ${\bf p}\cdot {\bf x}=wt-px$ and the integral is 
taken over the location of the 
insertion of one of the operators in the two-point function.
The thermal two-point function is given by
\be
{\cal G}(t,x)=\langle {\cal O}(t,x){\cal O}(0,0) \rangle_{T_H} = 
C \left({\pi T_+\over \sinh{\pi T_+ x_+}}\right)^{2h_+}
\left({\pi T_-\over \sinh{\pi T_- x_-}}\right)^{2h_+}.
\ee
Here $C$ is a normalization constant, unimportant for us,
and $x_{\pm}=t\pm x$ are the usual null coordinates. The
left and right temperatures are related to $T_H$ as:
\be
{2\over T_H}={1\over T_+}+{1\over T_-}.
\ee
The conformal dimension is, for a massive field of mass $m$
\be
h_+ = {1\over 2}(1+\sqrt{1+m^2}).
\ee
The integral (\ref{abs-prob-int}) gives 
\be\label{abs-prob}
\sinh{({\bf \beta}\cdot{\bf p})}\,
{(2\pi T_+)^{2h_+ -1} (2\pi T_-)^{2h_+ -1}\over \Gamma^2(2h_+)}
\left| \Gamma\left(h_+ + i{p_+ \beta_+\over 4\pi}\right)
\Gamma\left(h_+ + i{p_- \beta_-\over 4\pi}\right) \right|^2.
\ee
Here $\beta_{\pm} = 1/T_{\pm}$, and $p_{\pm}=\omega\pm p$.
This matches the absorption probability calculated from
the wave equation in the black hole background. For large
frequencies (large $\bf p$), the integral (\ref{abs-prob-int})
is dominated by $x_\pm \sim 1/p_\pm$. Thus, the large
frequency absorption probability is given by just the
two point function. Thus, our answer (\ref{em-prob})
is qualitatively correct, at least for large frequencies.

\subsection{Black hole creation by two point particles}
\label{sec:creation}

Let us now discuss the black hole creation process.
We will be schematic. A detailed analysis is presented in
the companion paper \cite{BH-Creation}.

The classical process of black hole creation by two point
particles was described by Matschull \cite{Matsch}. The
case of non-zero impact parameter is analyzed in \cite{Matsch-Rot}.
The case of massive particles is analyzed in \cite{Ivo-Formation}.

Let us consider the simplest case of a head-on collision of
two massive particles of equal mass $M/\pi=1-\alpha^2$. The elliptic
generators describing particles can be taken, for instance, to
be:
\be
A_1 = e^{s(\gamma_0+\kappa\gamma_2)}, \qquad
A_2 = e^{s(\gamma_0-\kappa\gamma_2)}.
\ee
Here $\kappa<1$ is the boost parameter and $s$ is related to the
identification angle via:
\be\label{mass-s}
{1\over 2}{\rm Tr}(A_{1,2}) = \cos{(s\sqrt{1-\kappa^2})}=\cos{\pi\alpha}.
\ee
The trace of the product
of these two generators plays the role of the order parameter
for the black hole creation. It is given by:
\be\label{order}
{1\over 2}{\rm Tr}(A_1 A_2) = 1- 2{\sin^2(\alpha)\over 1-\kappa^2}.
\ee
The same expression is obtained in \cite{Ivo-Formation}, the
only difference being that our $\alpha$ is twice that used in
there, and that we use a different parameterization of the boost
parameter. 

Massless particles correspond to the limit $\kappa\to 1$. In this 
limit the order parameter (\ref{order}) becomes $1-2s^2$. A black hole
is created when $s\geq 1$, see \cite{Matsch}. This is consistent
with one's intuition. Indeed, in order to create a black hole out
of massless particles the particles must be boosted to some minimal 
momentum. At this threshold a zero mass black hole is created.

Another important case to consider is that of two particles whose combined 
mass is enough to create a black hole for any value of the boost parameter.
As is not hard to see from (\ref{order}), this corresponds to 
$\alpha=\pi/2$. Then there is a black hole created for arbitrary small
values of the boost $\kappa$. 

Let us now indicate how all this can be described in the quantum theory.
As in the previous subsection, we would like to find a HH state whose
norm squared would give the probability for the process. The classical
process described above corresponds to creation of a single asymptotic 
region black hole. Thus, the relevant spatial slice geometry is that of
a disc with two operator insertions. The HH state is then, schematically
\be\label{bh-creation-ampl}
\Psi_{\rm HH}[a;\eta,\eta]=Z_{\rm CFT}\left[\,\,
\lower0.23in\hbox{\epsfig{figure=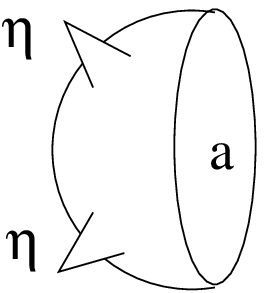,height=0.5in}}
\,\,\right].
\ee
This wave-function depends on the boundary condition $a$, on the
rest mass of the particles $\eta$, and on the relative position
of the insertion points, which encodes the particle's relative 
momentum. A precise relation between the position and momentum 
is given in the companion paper \cite{BH-Creation}. 

To obtain the BH creation probability let us take $|\Psi_{\rm HH}|^2$ 
and sum over all possible boundary conditions. Summing over boundary 
conditions is
equivalent to ``erasing'' them, and the result is the 4-point
function on the sphere:
\be
\sum_a |\Psi[a;\eta,\eta]|^2 = Z_{\rm LFT}\left[\,\,
\lower0.3in\hbox{\epsfig{figure=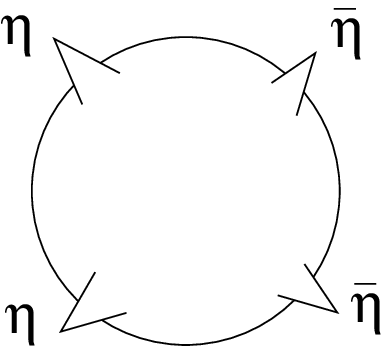,height=0.6in}}
\,\,\right].
\ee
This quantity has the interpretation of the probability of two
point particles colliding and forming a BH. To get 
the probability of creation of a {\it particular size} BH we
have to project the 4-point function on some intermediate
state. We have, schematically,
\be\label{prob}
Z_{\rm LFT}\left[\,\,
\lower0.3in\hbox{\epsfig{figure=4-point.eps,height=0.6in}}
\,\,\right] = \sum_P \,\,
\lower0.3in\hbox{\epsfig{figure=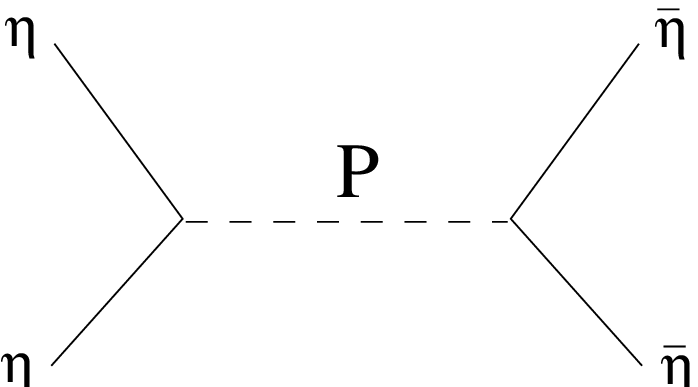,height=0.6in}}
\ee
The sum here is actually an integral, see \cite{BH-Creation}. 
Each term in the sum 
(\ref{prob}) has the interpretation of the probability of
creation of a BH of a particular size determined by the label $P$.

As is explained in \cite{BH-Creation}, in the semi-classical
limit of small AdS curvatures, the 4-point function is dominated
by the exponential of the classical Liouville action. Then, 
for a large BH created, the Liouville action is proportional
to the BH size. One thus obtains an exponentially small 
answer for the probability. As is explained in the companion paper
\cite{BH-Creation} the total probability of creating a horizon
is given by a sum of probabilities of all possible topologies
inside the horizon, the process (\ref{bh-creation-ampl}) giving
the simplest topology. It is argued in \cite{BH-Creation}
that the total probability is close to unity.

\section{Inner product and the S-matrix}
\label{sec:S-matrix}

In this section we introduce an inner product in the space
of HH states, and define the point particle S-matrix.
This leads to a stringy interpretation of our formalism. 

The state (\ref{state-1}) can be interpreted as the
amplitude for a black hole with three asymptotic regions
of size $a,b,c$. The state (\ref{state-1-part}) can be 
similarly interpreted as the amplitude for three
asymptotic region black hole with three point particles.
One can imagine a process in which one configuration goes into
the other. The amplitude for such a process should be given by
some inner product. There is a natural inner product that
can be constructed for HH states under consideration.
One should simply glue two Riemann surfaces with
boundary to form a closed surface, and consider the
full CFT partition function on it. To illustrate this,
let us consider two states. The first one $\Psi_1$ is the 
state of three asymptotic region black hole with three point 
particles. The second state $\Psi_2$ is that of the three asymptotic 
region black hole with a handle inside and two 
point particles:
\be
\Psi_1 = Z_{\rm CFT}\left[\,\,
\lower0.28in\hbox{\epsfig{figure=X1-part.eps,height=0.6in}}
\,\,\right], \qquad
\Psi_2 = Z_{\rm CFT}\left[\,\,
\lower0.28in\hbox{\epsfig{figure=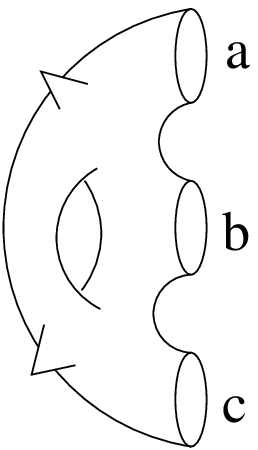,height=0.6in}}
\,\,\right].
\ee
Their inner product is defined as
\be\label{prod}
\langle \Psi_1 | \Psi_2 \rangle :=
 Z_{\rm CFT}\left[\,\,
\lower0.28in\hbox{\epsfig{figure=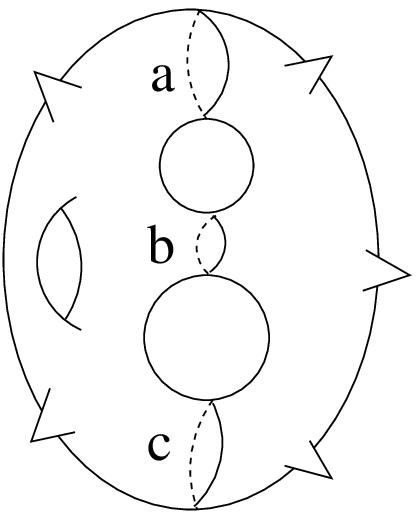,height=0.6in}}
\,\,\right].
\ee
It should be interpreted as the amplitude for the handle inside
the horizons plus two particles be converted into three
point particles. This process happens at fixed boundary
conditions for all three asymptotic regions. Of course,
this particular amplitude may be zero, but (\ref{prod})
illustrates the general principle. 

Let us now consider a different problem. Assume that one
is interested in the total amplitude for this process,
for all possible values of the boundary conditions at infinity.
This more general amplitude can be obtained by summing over
the boundary conditions. Such a sum over a complete set
is equivalent to having no boundary conditions at
all.\footnote{%
More precisely, one sums over all possible intermediate states, 
thus gluing two Riemann surfaces together. This is what we
mean by a sum over boundary conditions.} Thus, we have:
\be
\sum_{a,b,c} Z_{\rm CFT}\left[\,\,
\lower0.28in\hbox{\epsfig{figure=X3-part.eps,height=0.6in}}
\,\,\right] =
 Z_{\rm CFT}\left[\,\,
\lower0.28in\hbox{\epsfig{figure=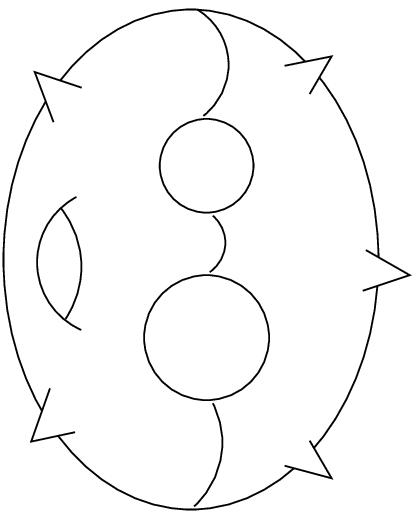,height=0.6in}}
\,\,\right].
\ee
This is the total amplitude for two particles to collide
with a handle and produce three particles. In this amplitude
one only fixes the mass of the particles and moduli
of the surface. One can consider an even more general amplitude
where only the masses are fixed. This is obtained by integrating
over the insertion positions, and over the surface moduli:
\be
\langle \alpha_1,\ldots,\alpha_k |
\alpha_{k+1},\ldots,\alpha_n \rangle_g = \int dm_X d\bar{m}_X\,\, 
Z_{\rm CFT}\left[\,\,
\lower0.17in\hbox{\epsfig{figure=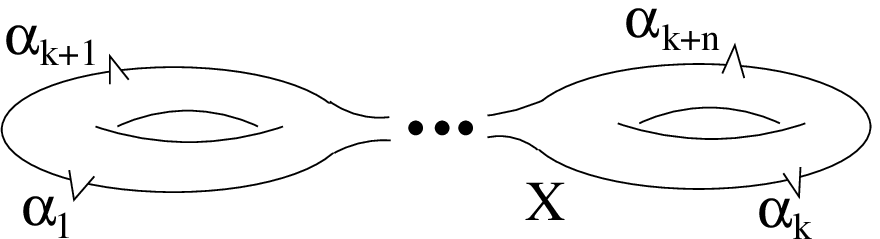,height=0.45in}}
\,\,\right].
\ee
Here $\alpha_1,\ldots,\alpha_k$ and 
$\alpha_{k+1},\ldots,\alpha_n$ are parameters of the incoming
and outgoing particles, $V_\alpha$ are the corresponding vertex operators.
This amplitude only depends on the genus of $X$ and on the
particle masses. To form the full S-matrix for the particles
one should in addition sum over the genus. Different genera
can be differently weighted by the ``string coupling constant'' $g_s$. 
The full S-matrix for particles is then given by the
following expression:
\be
\langle \alpha_1,\ldots,\alpha_k |
\alpha_{k+1},\ldots,\alpha_n \rangle =
{1\over g_s^2} \sum_g g_s^{2g} \int dm_X d\bar{m}_X\,\, 
Z_{\rm CFT}\left[\,\,
\lower0.17in\hbox{\epsfig{figure=s-matrix.eps,height=0.45in}}
\,\,\right].
\ee
This is the usual string theory expression for the S-matrix.
An important difference, however, is that the CFT we are working with
is non-critical. Our asymptotic states describing point particles are
also different from the usual asymptotic states of string theory.

\section{Remarks on unitarity and BH entropy}
\label{sec:remarks}

Up to this point we have assumed that the CFT in question is 
Liouville theory. As we shall now see, this is not quite right:
the ``correct'' CFT is a certain close relative of LFT that
incorporates the point particle states.

We start by reminding the reader that the Liouville
field theory is not capable of reproducing the entropy
of BTZ black holes. This important point was emphasized, e.g., 
in \cite{Carlip-What,Martinec}. This is because the effective central charge 
of Liouville field theory is equal to one, and, in Strominger's derivation 
\cite{Strom} of BTZ BH entropy, it is the effective central 
charge that has to be used in Cardy's formula for the density of 
states, see \cite{Carlip-What} for a discussion on this. 
This raised doubts, see \cite{Martinec}, on whether the quantum theory
of pure gravity in 2+1 dimensions is capable of reproducing 
the BH entropy microscopically. The string theory on AdS${}_3$,
on the other hand, does not suffer from this problem, because
the relevant CFT, which is the one coming from the D1/D5 system,
contains many fields, and its effective central charge is
large and equal to the Brown-Henneaux value. This seeming incapability
of pure gravity in 2+1 dimensions to account for the BH entropy
microscopically lead \cite{Martinec} to the conclusion that
``gravity is thermodynamics'' and one does not get anywhere by
attempting to quantize it: the correct microscopic theory is string theory. 
This is a viewpoint currently shared by most of the string community.

A related point is as follows. It can sometimes be heard that LFT is
non-unitary. This statement is not quite right as it stands, what is
actually meant by this is that LFT cannot contain non-normalizable
(conical singularity) states if one wants to have a unitary theory. 
This conclusion follows from the well-known fact that CFT
correlators contain only the normalizable states
from the continuous spectrum when decomposed into intermediate
states. This is true even for correlation functions of
vertex operators creating the non-normalizable (conical singularity)
states. Thus, allowing for the non-normalizable point particle
states one gets a non-unitary theory. Indeed, it is a theory in which one
has both normalizable and non-normalizable states
as external, but only normalizable states appear as the
internal states. The unitary LFT is a consistent truncation 
that only allows the normalizable states: both as external and
internal. The effective central charge of this unitary theory
is equal to one; it thus does not have enough states to
account for the BH entropy microscopically.

Thus, the only consistent theory seems to be the one with no 
particle (non-normalizable) states. There is something unsatisfactory 
with this picture, however. Indeed, LFT gives definite answers even for 
correlators of the non-normalizable states. These correlators definitely 
make sense, and are of great interest, for example, for the 
problem of accessory parameters for uniformization, and also for
understanding of the symplectic geometry of the moduli space
of punctured surfaces, see \cite{Takht-Elliptic}. It is
thus quite unfortunate if it is not possible to make sense
of LFT with non-normalizable states.

We have many reasons to believe that this is not so. In other words,
there are reasons to believe that there exists another CFT,
very closely related to the usual Liouville theory, which
consistently incorporates the LFT non-normalizable states. 
Semi-classically, this other CFT is related to the classical
geometry of Riemann surfaces, as is the usual LFT. However,
the correlation functions of this theory, when decomposed
into intermediate states, contain not only the states from
the continuum spectrum, but also the point particle
(non-normalizable) states. For example, the 4-point
function of point particles vertex operators contains,
in addition to the usual continuous part (\ref{prob}),
a part given by:
\be\label{interm-part}
\sum_{\eta'} Z_{\rm LFT}\left[\,\,
\lower0.25in\hbox{\epsfig{figure=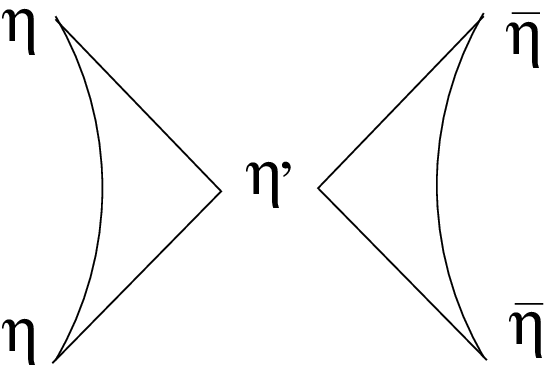,height=0.6in}}
\,\,\right].
\ee
What we have here is a product of two 3-point functions, and
the sum is taken over a set of intermediate particle states.
There are reasons to believe that the set of states that appears
is discrete. The quantization of the particle spectrum comes
as the requirement of rationality of the deficit angle.
It is expected that states with rational
deficit angle can be made normalizable by choosing the
inner product appropriately. We shall refer to the point
particle states that appear in (\ref{interm-part}) as
states of the discrete spectrum.

The reasons to believe the picture outlined are few. We shall
mention just a couple of them. First, it is known that the usual
LFT is intimately related to (essentially gives) the quantization
of the moduli space of compact Riemann surfaces. One can
expect a similar quantum theory to exist for the moduli
spaces of surfaces with conical singularities. The
quantization of this more general moduli space will
give conformal blocks of the CFT in question. A
reason to expect an appearance of the discrete part in
the spectrum is an analogy with the representation theory
of $\SL(2,\R)$. As is well-known, there are unitary
irreducible representations of two main types: the
continuous and discrete series. One can similarly
expect the appearance of these two series of representations
in the CFT. In fact, this is much more than an analogy,
for the theory of quantum Teichmuller spaces,
see \cite{Fock,Kashaev}, is based on the representation
theory of the quantum group $\SL_q(2,\R)$. The
usual Liouville theory is based on a certain
continuous series of representations of $\SL_q(2,\R)$,
see \cite{Teschner}. One can expect that there is another
CFT based on both the continuous and discrete series (continuous
representations possibly of a different series).

We have reasons to believe that the CFT whose structure we
outlined is a certain theory based on $\SL(2,\C)$ WZW model.
This theory was discussed in \cite{Teich}. It was shown in this
reference that it is this theory, not the usual LFT, that arises 
from the Euclidean path integral in $\Lambda<0$ gravity
in 3 dimensions. To
explain what that theory is, and what is its difference with the
usual LFT, we recall that there are two natural types of
structures that can be given a Riemann surface. These are:
(i) a conformal structure; (ii) a projective structure.
A conformal structure can be parametrized by prescribing a map
from the Riemann surface $X$ in question to the upper half-plane 
$\bf U$. Such a map is exactly what is used to obtain
the so-called operator formalism for LFT, see, e.g., 
\cite{Revisited}. The Liouville theory can be thought of as arising
by considering the path integral over such maps. A projective
structure, on the other hand, arises by giving a set of
charts of complex coordinates on $X$, with transition
functions given by fractional linear transformations.
In other words, a projective structure is a map from
$X$ to the complex plane. One can consider the path integral 
over such maps (projective structures). This defines a
theory that is different from, but is closely related to the 
usual LFT. We have reasons to believe
that this theory is essentially the LFT with point
particle states added to the spectrum. An attempt to demonstrate
this here would take us too far. We hope to return to this 
issue in the future.

Thus, in this and in the companion paper we have to ask the reader
to assume that there exists a CFT that is essentially LFT with
the point particle states added to the spectrum. This theory 
would solve the above-mentioned problem with BH entropy. Indeed,
this CFT would have zero
conformal dimension of the lowest lying state. Thus, its
effective central charge would be equal to the LFT value $c=1+6Q^2$. 
Therefore, at least naively, by adding point
particles to the theory one does get enough degrees of freedom
to account for the entropy. 

As is clear, for example, from (\ref{interm-part}), the role
of point particle states is to bring disconnected world-sheets
into the game. Thus, our proposal for obtaining a large
central charge CFT can be rephrased by saying that the central
charge comes via the disconnected world-sheets. We note that this
is essentially the same mechanism for obtaining a large
central charge as the one advocated in \cite{Ooghuri}.
It was argued in this reference that in the string theory
on AdS${}_3\times S^3\times T^4$ a large central charge
comes by considering the second quantized theory, which
includes disconnected world-sheets. This is very similar to the
mechanism proposed in this paper. This analogy at the very least
suggests a natural name for the CFT conjectured. We propose to call
it the second quantized Liouville field theory.

Let us summarize. The usual LFT does not account for the
BH entropy microscopically. However, one can conjecture 
the existence of another, closely related CFT, which
contains the point particle (non-normalizable) states
in the spectrum, and which thus must be capable of reproducing
the entropy. In the picture suggested the point particle states are
necessary to explain entropy microscopically. This is 
consistent with the fact that black holes can be made out of point
particles, and can evaporate into them. Thus, point particles
seem to be the constituents that black holes are made of. 

The CFT arising as the path integral over the projective structures
remains to be studied better, in order to demonstrate that it is indeed LFT
with the point particle states added to the spectrum.
However, there are many physically interesting situations in which
the particle states only appear as external. Then the usual Liouville theory
must be sufficient for calculations, at least in the semi-classical
regime when correlators are dominated by the classical Liouville action.
An example of a calculation along these lines is given in
the companion paper \cite{BH-Creation}. 

\section{Conclusions}

We have proposed a formalism in which amplitudes for physical
processes involving point particles can be computed as 
the quantum LFT partition function on Riemann surfaces. We have
sketched how this formalism can be applied to simple 
physical processes such as a particle emission by the
BTZ black hole, and the black hole creation process. A much
more detailed analysis of the BH production process is
presented in the companion paper \cite{BH-Creation}.

Our proposal to use Riemann
surfaces to do calculations in Lorentzian signature quantum
theory can be understood as a version of analytic continuation.
Indeed, it is only in the non-rotating case that one can
use the time symmetry surface to define HH states. When
there is no time symmetry plane, as is the case of spacetimes
with particles of non-zero momentum, or rotating spacetimes,
one defines HH states using the double $\tilde{X}$. The surface
$\tilde{X}$ naturally arises as the boundary of the Euclidean  
3-manifold that is the analytic continuation of the spacetime.

Our usage of Riemann surfaces leads us to the string theory
interpretation of the formalism. We have argued that
transition (scattering) amplitudes for point particles are
given by CFT correlators on closed Riemann surfaces. Sum
over all such amplitudes gives the S-matrix for particles.
This S-matrix is that of a non-critical string theory, whose
world-sheet CFT is essentially the Liouville field theory, see, however,
remarks in the previous section. Thus, this string
theory is essentially 2D gravity. Thus, somewhat 
surprisingly, we obtain that the hologram of the gravity
theory in 2+1 dimensions is also gravity, but in one dimension less. 
This is somewhat reminiscent of brane world ideas. An important difference, 
however, is that in the brane worlds scenarios the theory on the 
brane is always more than pure gravity: one has, in particular, 
gauge fields in addition to the metric.

Our construction have implications not just for the pure gravity
considered here, but also for other gravitational theories that include 
AdS${}_3$ as part of the background. Indeed, any such theory will
have non-trivial topology black holes and point particles as
its classical solutions. Since topologically non-trivial black holes 
seem to force one to consider higher genus Riemann surfaces, 
this leads to holographic theory being a string theory.
These ideas applied to the usual string theory AdS${}_3$/CFT
setting lead to a puzzle, for they seem to suggest that one string 
theory in bulk is dual to another string theory on the boundary.
It would be of interest to find whether this interpretation is
correct, and, in case it is, obtain a relation between the two
string theories. 

\noindent
{\large \bf Acknowledgments}

I would like to thank J.\ Baez, L.\ Freidel, J.\ Hartle, G.\ Horowitz,
R.\ Myers and J.\ Teschner for discussions. The author was supported by
the NSF grant PHY00-70895.

\newcommand{\hep}[1]{{\tt hep-th/{#1}}}
\newcommand{\gr}[1]{{\tt gr-qc/{#1}}}

\end{document}